\documentclass[twocolumn,floats,floatfix,superscriptaddress,aps,pra]{revtex4}
\usepackage{subfigure}
\usepackage{psfrag,graphicx}
\usepackage{bm,color}
\usepackage{latexsym}
\usepackage{epstopdf}
\usepackage[english]{babel}
\usepackage{amsmath,amssymb,amsfonts}
\usepackage{appendix}
\usepackage{bbold}
\usepackage[dvipsnames]{xcolor}

\definecolor{mygrey}{gray}{0.35}
\definecolor{myblue}{rgb}{0.2,0.2,0.8}
\definecolor{myzard}{cmyk}{0,0,0.05,0}
\definecolor{mywhite}{rgb}{1,1,1}
\definecolor{mywhite}{rgb}{1,1,1}
\definecolor{myred}{rgb}{1,0.,0.3}

\def\be{ \begin{equation}}
\def\ee{ \end{equation}}
\def\bse{  \begin{subequations}}
\def\ese{  \end{subequations}}
\def\bea#1\ea{\begin{align}#1\end{align}}

\def\bi{\begin{itemize}}
\def\ei{\end{itemize}}

\def\bt{\begin{tabular}}
\def\et{\end{tabular}}

\def\ket#1{\vert #1 \rangle}

\def\half{\tfrac12}

\def\3half{\tfrac32}

\def\to{\rightarrow}

\def\B{\mathbf{B}}
\def\U{\mathbf{U}}
\def\C{\mathbf{C}}
\def\H{\mathbf{H}}

\def\C{\mathbf{C}}

\def\P{\mathbf{P}}

\def\R{\mathbf{R}}

\DeclareMathOperator*{\SumInt}{%
\mathchoice%
  {\ooalign{$\displaystyle\sum$\cr\hidewidth$\displaystyle\int$\hidewidth\cr}}
  {\ooalign{\raisebox{.14\height}{\scalebox{.7}{$\textstyle\sum$}}\cr\hidewidth$\textstyle\int$\hidewidth\cr}}
  {\ooalign{\raisebox{.2\height}{\scalebox{.6}{$\scriptstyle\sum$}}\cr$\scriptstyle\int$\cr}}
  {\ooalign{\raisebox{.2\height}{\scalebox{.6}{$\scriptstyle\sum$}}\cr$\scriptstyle\int$\cr}}
}

\begin{document}

\begin{abstract}
Laser-induced-continuum-structure(LICS) allows for coherent control techniques to be applied in a Raman type system with an intermediate continuum state.
The standard LICS problem involves two bound states coupled to one or more continua.
In this paper we discuss the simplest non-trivial multistate generalization of LICS which couples two bound levels, each composed of two degenerate states through a
common continuum state.
We reduce the complexity of the system by switching to a rotated basis of the bound states, in which different sub-systems of lower dimension evolve
independently.
We derive the trapping condition and explore the dynamics of the sub-systems under different initial conditions.
\end{abstract}

\author{K. N. Zlatanov}
\affiliation{Department of Physics, Sofia University, James Bourchier 5 blvd, 1164 Sofia,
Bulgaria}
\affiliation{Institute of Solid State Physics, Bulgarian Academy of Sciences, Tsarigradsko chaussée 72, 1784 Sofia, Bulgaria}
\author{N. V. Vitanov}
\affiliation{Department of Physics, Sofia University, James Bourchier 5 blvd, 1164 Sofia,
Bulgaria}
\title{Multilevel Laser Induced Continuum Structure}
\date{\today }
\maketitle


\section{Introduction}

Coherent manipulation of quantum states is at the core of contemporary quantum physics.
The development of analytical models \cite{Rabi1937,Landau1932,Zener1932,Stuckelberg1932,Majorana1932,Rosen1932} treating laser interaction between two,
three and more levels \cite{Allen1975,Shore1990} is fundamental for the understanding and the demonstration of effects like rapid adiabatic passage (RAP)
\cite{Vitanov2001RAP}, stimulated Raman adiabatic passage (STIRAP) \cite{Vitanov2017STIRAP} and electromagnetically induced transparency (EIT)
\cite{Harris1997,Fleischhauer2005,Ullah2018}, to name just a few.
Generally, the problems of coherent population transfer and state preparation involve only bound states that lye deep inside the potential of the atom, which have
well defined discrete energies.
There exist, however, unbound states lying outside the potential of the atom distributed continuously in energy, and therefore termed continuum states.

The simplest, flat continuum possesses no structures.
However, structures in the continuum can be induced by laser fields.
The emergence of resonance structures in the continuum has been described by Fano in his seminal paper \cite{Fano1961} on autoionisation.
If a bound state is coupled to the continuum by a strong laser field, the latter ``embeds'' this state into the continuum.
Scanning through this energy range by another (weak) laser field which couples a second state to the same continuum reveals a resonance structure
known as \emph{laser-induced-continuum-structure (LICS)} \cite{Armstrong1975}.
A suitable choice of the two-photon detuning between the two bound states, given by the \emph{trapping condition} \cite{Coleman1982}, allows to suppress the
ionization in theory to zero.
In experiments, ionization suppression by as much as 80\% has been achieved by Halfmann and co-workers \cite{Halfmann1998, Yatsenko1999}.

When the trapping condition is fulfilled, coherent processes between the bound states become possible.
The most prominent among these processes is coherent population transfer between the bound states.
In particular, the counterintuitive arrangement of the light fields, as in STIRAP, allows for population transfer between the bound states through the continuum state,
with little or even no ionization \cite{Carroll1992,Carroll1993,Carroll1995,Carroll1996}.
In experiments, population transfer efficiency of about 20\% has been achieved \cite{Peters2005,Peters2007}.
Optimal population transfer occurs only if the trapping condition on the two-photon detuning is achieved, especially if it is satisfied during the entire evolution of the
system \cite{Nakajima1994}.
Because the driving fields are time-dependent the trapping condition becomes time-dependent too.
This requirement can be met to some extent by techniques like pulse chirping \cite{Paspalakis1997,Paspalakis1998,Vitanov1997} and non-ionising Stark shifts \cite{Yatsenko1997,Vitanov1997}.
Incoherent decay channels can be suppressed by coupling of additional states to the system \cite{Unanyan1998}.
Viewed from the opposite angle, LICS can be used aslo to increase ionisation via STIRAP into a continuum \cite{Rangelov2007}.

The standard LICS problem, reviewed by Knight \cite{Knight1990}, describes a Raman type transition between two bound states with intermediate common
continuum state.
Further theoretical development expands the model by including multiple continua \cite{Bohmer2002}, while keeping two bound states.
Three bound states coupled to a continuum have been studied too \cite{Unanyan2000}, and in this system two trapping conditions have been derived.
An attempt to include sub-level structure of the bound states was realised by switching to the Laplace domain \cite{Parzynski1987,Parzynski1988}, which has the
limitation of specific excitation patterns.
In addition, simply adding more bound states to the standard LICS model drastically increases the complexity and prevents the derivation of a trapping condition
because, in order to find such, one has to solve a characteristic polynomial equation of growing order as the number of states increases.

Some experimental demonstrations of LICS revealed interesting effects about chemical reactions \cite{Shnitman1996,Thanopulos2010}, ionisation branching
\cite{Eramo1998} and harmonic generation \cite{Pavlov1982}.
Further experiments in molecules \cite{Faucher1999} showed how vibrational states can influence interference patterns and emphasised the need of more
complicated LICS models, that account for molecular properties.
The high-order LICS effects in complex systems remain largely unstudied due to the lack of more sophisticated theoretical models, that account for multiple bound
states.

In this paper we treat the excitation dynamics of degenerate levels of ground and excited bound states, coupled to a common continuum.
In order to address problems with the growing complexity of a large number of bound states  we show how a multilevel LICS system can be reduced to independent
sub-systems of smaller dimension, by a proper change of basis.
We further derive the trapping condition for the population transfer between the bound states of the sub-systems.
We explore the dynamics for different initial conditions and their Fano profiles, and we suggest new applications of LICS based on our findings.

This paper is organized as follows.
In Section~\ref{Sec:two} we introduce the problem.
Section ~\ref{Sec:results} treats population trapping, different initializations of the system and the associated Fano profiles.
We conclude our presentation in Section~\ref{Sec:conclusion}.


\section{Multilevel LICS system}\label{Sec:two}


\begin{figure}[tb]
\bt{c}
\includegraphics[width=0.45\columnwidth]{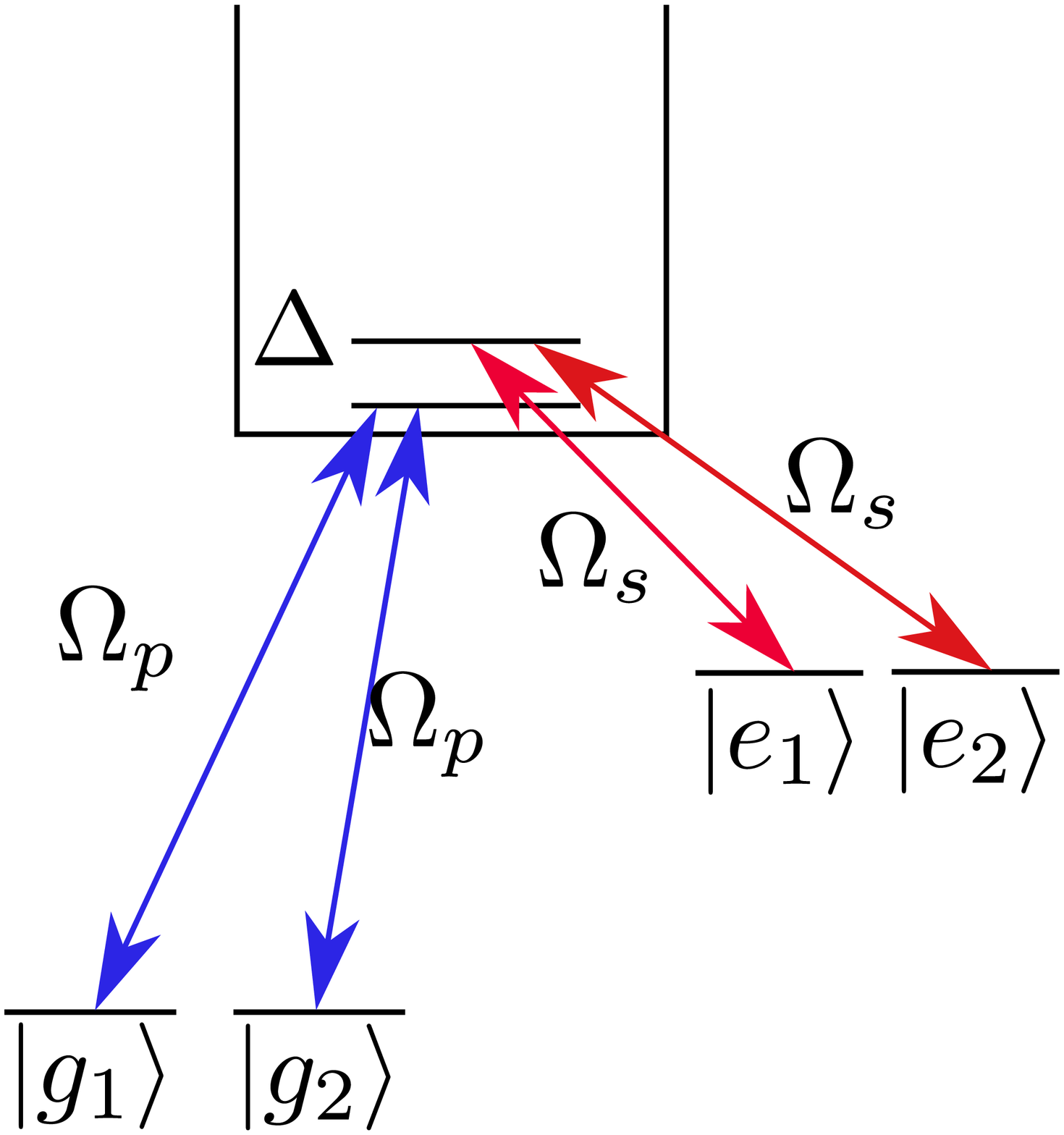}\llap{
  \parbox[b]{3.3 in}{(a)\\\rule{0ex}{1.61in}
  }} \\ \includegraphics[width=0.390\columnwidth]{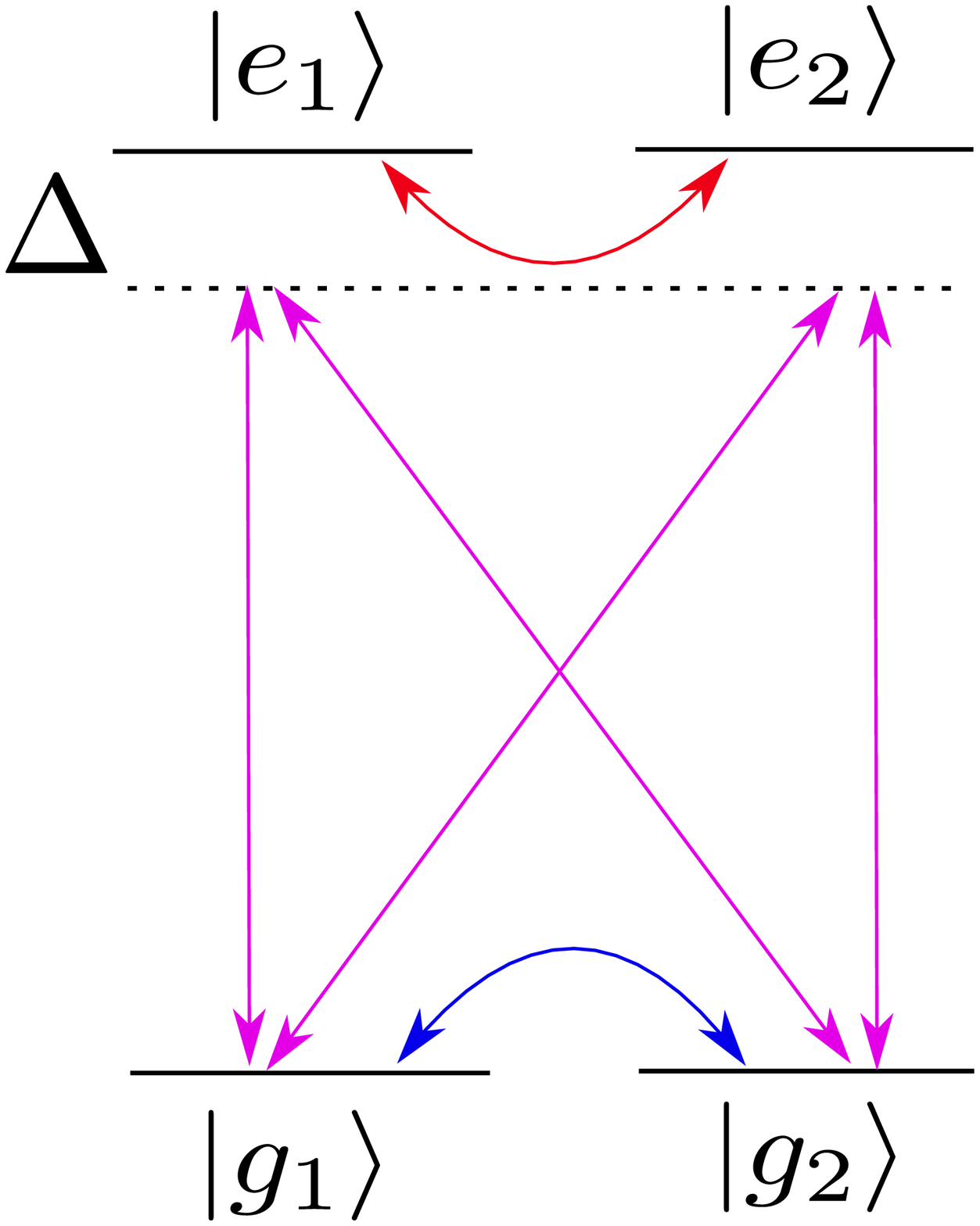}\llap{
  \parbox[b]{3.0 in}{(b)\\\rule{0ex}{1.61in}
  }}
\et
\caption{(Color online)
Multilevel LICS coupling scheme.
(a) Four bound states coupled to common continuum.
(b) The effectively reduced system of inter-coupled bound states, by elimination of the continuum.
}\label{fig:System}
\end{figure}

We consider the simplest multilevel system, that has more than one state in each ground and excited bound levels, namely two ground and two excited states coupled
to a common continuum by two independent lasers as illustrated in Fig.~\ref{fig:System} (a).
For the sake of simplicity we shall ignore highly detuned transitions, continuum-continuum transitions and we will assume the ground and excited levels to be
degenerate.

The evolution of the system is governed by the time dependent Schrodinger equation($\hbar=1$)
\be
i\frac{d}{dt}\mathbf{A}(t)=\mathbf{H}_{Sch}(t)\mathbf{A}(t),
\ee%
driven by a Hamiltonian which in Schrodinger picture reads
\be
\H_{Sch}=\left[
\begin{array}{ccccc}
 \omega _{g_1} & 0 & 0 & 0 & -\hat{\Omega }_{g_1,p} \\
 0 & \omega _{g_2} & 0 & 0 & -\hat{\Omega }_{g_2,p} \\
 0 & 0 & \omega _{e_1} & 0 & -\hat{\Omega }_{e_1,s} \\
 0 & 0 & 0 & \omega _{e_2} & -\hat{\Omega }_{e_2,s} \\
 -\hat{\Omega }_{g_1,p} & -\hat{\Omega }_{g_2,p} & -\hat{\Omega }_{e_1,s} & -\hat{\Omega }_{e_2,s} & \omega _{\epsilon _c}
   \\
\end{array}
\right].
\ee
The operator Rabi frequencies acting on the state vector's component $a_n(t)$ are defined as
\be
\hat{\Omega }_{i,l}a_n(t)= \int_{0}^{\infty} \Omega_{i,\epsilon,l}(t) \cos(\omega_l t)a_n(t) d\epsilon,
\ee
linking the transition between the $i$-th bound state and a continuum state of energy $\epsilon,$ due to the $l$-th laser.
We can eliminate the continuum state by the procedure given in \cite{Knight1990} and also switch the phase picture by $a_n(t) \to c_n(t) =a_n(t)\exp(i \omega_n t)$.
Thus we reduce the system to the effective interaction among the bound states illustrated in Fig.~\ref{fig:System} (b) whose evolution is given by
\be
i\frac{d}{dt}\mathbf{C}(t)=\mathbf{H}(t)\mathbf{C}(t),  \label{Sch}
\ee%
The Hamiltonian driving the reduced system is a non-Hermitian matrix given by
\be
\mathbf{H}(t)=-\frac{1}{2}(\H_0+i\H_1),\label{H}
\ee
with
\bse
\bea
\H_0&=\left[
\begin{array}{cccc}
 -2 \text{$\delta $S}_g & q_{\text{gg}} \Gamma _{g} & q_{\text{eg}} \Gamma _{\text{eg}} & q_{\text{eg}} \Gamma _{\text{eg}} \\
 q_{\text{gg}} \Gamma _{g} & -2 \text{$\delta $S}_g & q_{\text{eg}} \Gamma _{\text{eg}} & q_{\text{eg}} \Gamma _{\text{eg}} \\
 q_{\text{eg}} \Gamma _{\text{eg}} & q_{\text{eg}} \Gamma _{\text{eg}} & -2 \left(\Delta +\text{$\delta   $S}_e\right) & q_{\text{ee}} \Gamma _{e} \\
 q_{\text{eg}} \Gamma _{\text{eg}} & q_{\text{eg}} \Gamma _{\text{eg}} & q_{\text{ee}} \Gamma _{e} & -2  \left(\Delta +\text{$\delta $S}_e\right)
\end{array}
\right],\label{H0}\\
\H_1&=\left[
\begin{array}{cccc}
 \Gamma _g & \Gamma _{g} & \Gamma _{\text{eg}} & \Gamma _{\text{eg}} \\
 \Gamma _{g} & \Gamma _g & \Gamma _{\text{eg}} & \Gamma _{\text{eg}} \\
 \Gamma _{\text{eg}} & \Gamma _{\text{eg}} & \Gamma _e & \Gamma_{e} \\
 \Gamma _{\text{eg}} & \Gamma _{\text{eg}} & \Gamma _{e} & \Gamma _e
\end{array}
\right] . \label{H1}
\ea\label{H0H1}
\ese
The diagonal elements in Eqs.~\eqref{H0H1} are defined as follows.
First of all,
\be
\Delta=E_{e_i}-E_{g_i}+\omega_{s}-\omega_{p}
\ee
is the reduced two-photon detuning connecting the ground and excited states through the continuum.
Since we assume degeneracy, $\Delta$ is the same for both bound levels. The ionization rate due to a single laser is given by
\be
\Gamma _{k} = \frac{1}{2} \pi    \left|\Omega _{kc_{\epsilon},l}\right|^2,\label{Gama_sl}
\ee
where $k$ runs over the bound states $g$ and $e$, $l$ runs over the lasers, and  $c_{\epsilon}$ designates a continuum state with energy $\epsilon$ in the new
phase picture.
The Stark shifts caused by the lasers are defined as
\be
\delta S_k= -\mathcal{P.V} \displaystyle\SumInt d\epsilon\frac{\left|\Omega _{kc_{\epsilon},l}\right|^2}{4 (\epsilon-E_k-\omega_l)}.
\ee

The off-diagonal elements
\be
\Gamma_{ij}=\frac{1}{2} \pi    \Omega _{ic_{\epsilon},a} \Omega _{jc_{\epsilon},b}=\sqrt{\Gamma_i\Gamma_j},\quad i\neq j
\ee
give the two-photon coupling between the bound ground  $i$ and excited $j$ states through the continuum state $c_{\epsilon}$ due to the interaction with both
lasers. 

The continuum affects the evolution of the system by the Fano parameters
\be
q_{ij}= \frac{\mathcal{P.V.} \displaystyle\SumInt d\epsilon\frac{\Omega _{i c_\epsilon ,l} \Omega _{j c_\epsilon,m}^*}{2 (\epsilon-E_g-\omega_l)}}{\Gamma_{ij}},
\ee
where $\omega_l$ is the frequency of each laser $l$ which drives the respective bound $\to$ continuum transition, while $m$ indicates the  laser driving the
continuum $\to$ bound transition.
For example, $l=p=m$ for $g\to c_{\epsilon}\to g$ transition, and $l=p$, $m=s$ for $g\to c_{\epsilon}\to e$ transitions.
Since we consider a degenerate system, we can distinguish between three different Fano parameters, namely (i) $q_{gg}$ for transitions linking $c_{g_1}$ and
$c_{g_2}$ through the continuum, (ii) $g_{ee}$ for transitions linking $c_{e_1}$ and $c_{e_2}$, and (iii) transitions between bound states through the continuum
$c_{g_i}$ and $c_{e_j}$.	

The elimination of the continuum state reduces the problem to four inter-coupled bound states.
Although this elimination simplifies the system, it does not amount to solving the problem.
For example, if we try to derive a population trapping condition by imposing conditions on the characteristic polynomial of Eq.~\eqref{H}, as in \cite{Knight1990}, we
have to solve a forth-order equation for the eigenvalues of the system.
If we add more states to the system the problem becomes algebraically unsolvable in this basis.

A way out of this difficulty is the similarity transformation
 \be
\U=\left[
\begin{array}{cc}
 \R(\theta) & \mathbf{0}\\
  \mathbf{0} & \R(\theta)
\end{array}
\right],\label{U_ms}
\ee
where the matrix $\R$ reads
\be
\R=\left[
\begin{array}{cc}
 \cos (\theta ) & \sin (\theta ) \\
 -\sin (\theta ) & \cos (\theta ) \\
\end{array}
\right].
\ee
Thus we can reduce the system to independent sub-systems, and allow for a simpler derivation of the trapping condition by fixing the rotation angle at $\theta=\pi/4,$
and further applying the shift transformation
\be
\P=\left[\begin{array}{cccc}
 1 & 0 & 0 & 0 \\
 0 & 0 & 1 & 0 \\
 0 & 1 & 0 & 0 \\
 0 & 0 & 0 & 1 \\
\end{array}\right].
\ee
The transformed Hamiltonian then reads
\bea
\widetilde{\H}&=\P\U\H\U^{\dagger}\P
=\left[\begin{array}{cc}
 \H_b & \mathbf{0} \\
 \mathbf{0} & \H_d \\
\end{array}\right],
\ea
where the matrices $\H_b$ and $\H_d$ are given by
\bse
\bea
\H_b &=\left[
\begin{array}{cc}
 \text{$\delta $S}_g-\frac{1}{2} \left(q_{\text{gg}}+2 i\right) \Gamma _g & -\left(q_{\text{eg}}+i\right)
   \sqrt{\Gamma _e \Gamma _g} \\
 -\left(q_{\text{eg}}+i\right) \sqrt{\Gamma _e \Gamma _g} & \Delta -\frac{1}{2} \left(q_{\text{ee}}+2
   i\right) \Gamma _e+\text{$\delta $S}_e \\
\end{array}
\right],\\
\H_d &=\left[
\begin{array}{cc}
 \frac{q_{\text{gg}} \Gamma _g}{2}+\text{$\delta $S}_g & 0 \\
 0 & \Delta +\frac{q_{\text{ee}} \Gamma _e}{2}+\text{$\delta $S}_e \\
\end{array}
\right].
\ea
\ese

The combined transformation upon the state vector reads,
\be
\P\U\C(t)=
\frac{1}{\sqrt{2}}\left[
\begin{array}{c}
 c_{g_1}+c_{g_2} \\
 c_{e_1}+c_{e_2} \\
 c_{g_2}-c_{g_1} \\
 c_{e_2}-c_{e_1} \\
\end{array}
\right]=\left[
\begin{array}{c}
 b_{g} \\
 b_{e} \\
 d_{g} \\
 d_{e} \\
\end{array}
\right].
\ee
In the next section, we investigate the behaviour of the reduced system, and derive the conditions for population trapping.

\section{Excitation probabilities and Fano profiles}\label{Sec:results}

The benefit of the block-diagonalization of Eq.~\eqref{H} is that now $\H_b$ and $\H_d$ operate independently on the superposition states $\{b_g,b_e\}$ and
$\{d_g,d_e\}$, respectively.
This allows us to derive separate trapping conditions.
In our current example this is solely for $\H_b$ since $\H_d$ is composed of "dark" states, which do not participate in the excitation.
In order to find the trapping condition, we require to have a real eigenvalue of $\H_b$, i.e. a real solution of its characteristic polynomial.
Thus we find the trapping condition to be
\be
\Delta=\half\left(\Gamma _e q_{\text{ee}}-\Gamma _g q_{\text{gg}}\right) +q_{\text{eg}} \left(\Gamma _g-\Gamma
   _e\right)+\text{$\delta $S}_g-\text{$\delta $S}_e.\label{Trapping}
\ee
 The main difference between the trapping condition of a standard two-level LICS and our model is the additional term deriving from the couplings of the same bound
 level through the continuum.
 In order to explore this difference we turn to the solution of the system driven by $\H_b.$
 For the sake of simplicity, we look at continuous-wave (cw) excitation where the two lasers act for a specific time period $T$.
 If pulsed excitation is used the time dependence for the couplings and the detuning has to be the same \cite{Nakajima1994}, which is not a particular difficulty for
 modern laser systems.

\subsubsection{System initialized in a coherent superposition of states}

In the simplest scenario, we initialize the system in a coherent superposition of the ground states $[c_{g_1}(0)+c_{g_2}(0)] / {\sqrt{2}} = b_g(0) = 1$.
This initial condition ensures that the dark states will not be populated and the evolution of the system will be governed by
\be
i \frac{d}{dt}\B=\H_b\B,  \label{HuShr}
\ee%
with $\B=[b_{g}, b_{e}]^T$.

The general solution of Eq.~\eqref{HuShr} is too cumbersome to be presented here even for cw excitation.
A simplified solution can be generated if we substitute the trapping condition of Eq.~\eqref{Trapping} back into the solution which then reads
\bse
\bea
b_g&=\frac{\left[\Gamma _e+\Gamma _g e^{i t \left(q_{\text{eg}}+i\right) \left(\Gamma _e+\Gamma   _g\right)}\right] e^{-\frac{1}{2} i t \left(\Gamma _g \left(2
q_{\text{eg}}-q_{\text{gg}}\right)+2   \text{$\delta $S}_g\right)}}{\Gamma _e+\Gamma _g},\\
b_e&=\frac{\sqrt{\Gamma _e \Gamma _g} \left[ e^{i t \left(q_{\text{eg}}+i\right) \left(\Gamma _e+\Gamma   _g\right)}-1\right] e^{-\frac{1}{2} i t \left(\Gamma _g
\left(2 q_{\text{eg}}-q_{\text{gg}}\right)+2   \text{$\delta $S}_g\right)}}{\Gamma _e+\Gamma _g}.
\ea
\ese
\begin{figure}[tb]
 \includegraphics[width=0.85\columnwidth]{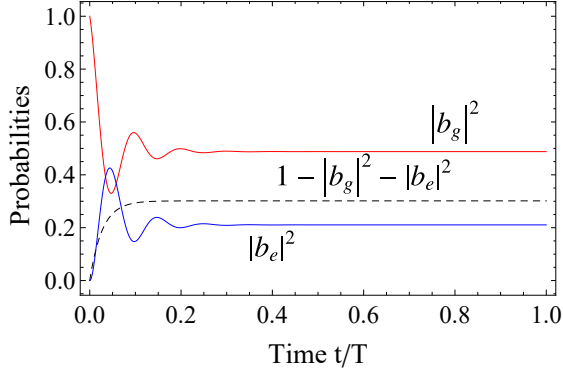}
\caption{(Color online)
Probability amplitudes versus time $t/T$ normalized over the duration of the excitation for system initialized in the "bright" ground state.
The red and blue lines give the ground and excited bright states, respectively, while the dashed black line depicts the  ionization probability.
\\
The excitation parameters are set to $\delta S_g=0.5 T^{-1} $, $\delta S_e=0.6 T^{-1} $, $\Gamma _g=5.5 T^{-1},$ $\Gamma _e=12.74 T^{-1},$ $q_{gg}=2.3$,
$q_{eg}=3.4$, $q_{ee}=5$.
}\label{fig:P for cs}
\end{figure}

The evolution of the system under the trapping condition is shown in Fig.~\ref{fig:P for cs}.

\subsubsection{System initialized in one of the ground states}

\begin{figure}[tb]
 \includegraphics[width=0.85\columnwidth]{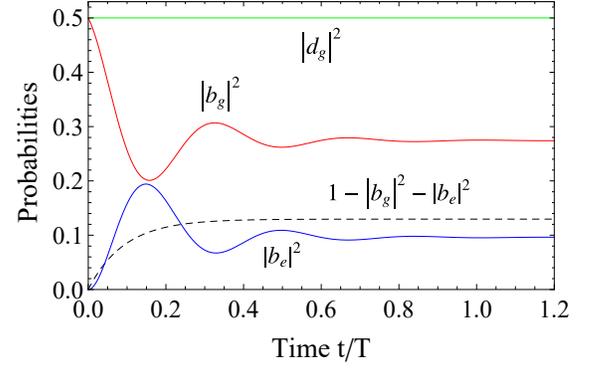}
\caption{(Color online)
The same as Fig.~\ref{fig:P for cs} but for the system initially in state $\ket{g_1}$.
 The dashed black line depicts the  ionization probability, while the green line gives the population in
the dark state. The excitation parameters are set to $\delta S_g=0.5 T^{-1} $, $\delta S_e=0.6 T^{-1} $, $\Gamma _g=5.5 T^{-1},$ $\Gamma _e=12.74 T^{-1},$
$q_{gg}=2.3$, $q_{eg}=3.4$, $q_{ee}=5$.
}\label{fig:P for g1}
\end{figure}

The multilevel LICS problem is quite sensitive to the initial condition.
If we instead initialize the system in one of its ground states, say $c_{g_1}(0)=1$, we will have to account for one of the dark states, namely $d_g$.
The evolution of the system is then governed by
\be
i \frac{d}{dt}\left[
\begin{array}{c}
 \B \\
 d_{g}
\end{array}
\right]=  \left[
\begin{array}{cc}
 \H_b & 0 \\
 0 & \frac{q_{\text{gg}} \Gamma _g}{2}+\text{$\delta $S}_g
\end{array}
\right] \left[
\begin{array}{c}
\B \\
 d_{g}
\end{array}
\right]\label{G1state}.
\ee
Due to the block-diagonal form of Eq.(\ref{G1state}) the solution for $\B$ remains unchanged besides a factor accounting for the new initial condition.
The solution of Eq.~\eqref{G1state} with imposed trapping condition and accounting for the dark state reads
\bse
\bea
b_g&=\frac{\left[\Gamma _e+\Gamma _g e^{i t \left(q_{\text{eg}}+i\right) \left(\Gamma _e+\Gamma _g\right)}\right] e^{-\frac{1}{2}
   i t \left[\Gamma _g \left(2 q_{\text{eg}}-q_{\text{gg}}\right)+2 \text{$\delta $S}_g\right]}}{\sqrt{2} \left(\Gamma   _e+\Gamma _g\right)},\\
b_e&=\frac{\sqrt{\Gamma _e \Gamma _g} \left[e^{i t \left(q_{\text{eg}}+i\right) \left(\Gamma _e+\Gamma _g\right)}-1\right]
   e^{-\frac{1}{2} i t \left[\Gamma _g \left(2 q_{\text{eg}}-q_{\text{gg}}\right)+2 \text{$\delta $S}_g\right]}}{\sqrt{2}
   \left(\Gamma _e+\Gamma _g\right)},\\
d_g&=-\frac{e^{-\frac{1}{2} i t \left(2 \text{$\delta $S}_g+\Gamma _g q_{\text{gg}}\right)}}{\sqrt{2}}.
\ea
\ese
The population evolution is depicted in Fig.~\ref{fig:P for g1}.
We note the difference in the ionisation as well as the diminishing excitation of the bright states.
This is the direct consequence of the initial condition.
With the initial condition $c_{g_1}(0)=1$ the ionisation
\be
I=1-\left|b_g\right|^2-\left|b_e \right|^2-\left|d_g\right|^2\label{ionisation g1}
\ee
decreases, since the dark state tends to preserve half of the population among $c_{g_1}$ and $c_{g_2}.$
Consequently, less population can pass through the continuum to the excited bound states.
This behaviour outlines the importance of initializing the system at $b_g(0)=1,$ since in that case the term $|d_g|^2$ in Eq.~\eqref{ionisation g1} vanishes.

\subsubsection{Fano profile}

Finally we point out significant differences between the Fano ionisation profiles of our four-level model and the two-level model of \cite{Knight1990}, whose evolution
is driven by
\be
\H_{2lvl}\left[
\begin{array}{cc}
 \text{$\delta $S}_g-\frac{i \Gamma _g}{2} & -\frac{1}{2} \left(q_{\text{eg}}+i\right) \sqrt{\Gamma _e \Gamma _g} \\
 -\frac{1}{2} \left(q_{\text{eg}}+i\right) \sqrt{\Gamma _e \Gamma _g} & \Delta -\frac{i \Gamma _e}{2}+\text{$\delta $S}_e \\
\end{array}
\right].\label{H2lvl}
\ee
The structure of $\H_b$ and $\H_{2lvl}$ is very similar. The excitation differs by a factor of $\half$ and the diagonal elements are effectively shifted, so naturally one
can expect the same Fano profile, whose minimum is also shifted.
\begin{figure}[tb]
 \includegraphics[width=0.85\columnwidth]{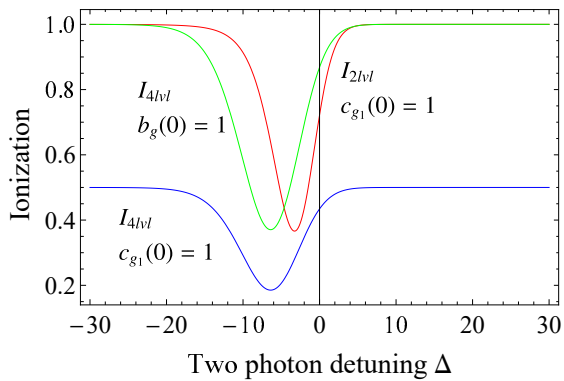}
\caption{(Color online)
Ionisation versus two photon detuning for different  system configurations.
The red line shows the ionisation for the standard two-state LICS \cite{Knight1990}.
The  blue and green lines show the ionisation for the four-state model of Eq.~\eqref{ionisation g1} with the population initially respectively in state $\ket{g_1}$ and the
bright state $\ket{b_g}$.
The excitation parameters are set to $\delta S_g=0.5T^{-1} $, $\delta S_e=0.6T^{-1} $, $\Gamma _g=5.5T^{-1},$ $\Gamma _e=12.74T^{-1},$ $q_{gg}=2.3$,
$q_{eg}=3.4$, $q_{ee}=5$, $t/T=6$.
}
\label{fig:Fano}
\end{figure}
In Fig.~\ref{fig:Fano} we show the ionisation profiles for different models and different initial conditions.
As we see in the figure, the profile of the two-level system is similar to the the profile of the four-level system initialised in a bright state.
This is not the case for a system initialised in the $\ket{g_1}$ ground states. Looking at the ionisation when both the two-state and four-state models are initialized in
state $\ket{g_1}$, we find that the minimum of the ionisation of the two-state model can correspond to a significant ionisation of the four-state model. This mismatch
shows that approximations ignoring the presence of nearby laying states, streaming for example from magnetic quantum number, do not predict the correct Fano
resonance.
Due to the dark state the ionization of the system can not exceed $\frac12,$ since it keeps half of the population inside the bound states.
This last feature is quite significant and can be harnessed in a few useful ways.
For example, if we want to break a chemical bond which is surrounded by multiple states it is best if we first create a coherent superposition and then ionise at a
maximum of the Fano profile.
Alternatively, by comparing Fano profiles we can judge about the structure of the system, since the more dark states are involved, the smaller the ionisation, as each
dark state will tend to keep more population bounded.

Finally we want to point out that a fulfilled trapping condition does not mean vanishing ionisation but rather a minimum. It ensures that one of the eigenvalues of the
Hamiltonian will be real, but not all. Thus a decay channel is open through the states with complex eigenvalues.

\subsubsection{Non-degeneracy}

Hitherto we have assumed that the states in each level are degenerate.
At first glance, it appears as a strong condition on the system not only because real systems are non-degenerate but it also equalises the strength of all
bound-continuum-bound transitions, as well as the Stark shifts and the Fano parameters.
However such differences will be of the order of the energy shifts among the states in the bound levels and are also quite small compared to the ionisation couplings
$\Gamma_{ij}$.
In order to estimate under what circumstances we can treat the system as degenerate we investigate numerically the evolution of the non-degenerate system driven
by the Hamiltonian
\be
\mathbf{H}_{nd}(t)=
 \left[ \begin{array}{cc}
 \mathbf{\Delta}_{g} & \mathbf{\Gamma}\\
  \mathbf{\Gamma} &  \mathbf{\Delta}_{e}
\end{array} \right], \label{H_nd}
\ee
where $\mathbf{\Gamma}$ are the coupling matrices of the degenerate Hamiltonian of Eq.~\eqref{H0H1} and the modified ground and excited blocks lye on the
diagonal,
\bse
\be
 \mathbf{\Delta}_{g} =
 \left[ \begin{array}{cc}
 \text{$\delta $S}_g-\frac{i \Gamma _g}{2} & -\frac{1}{2} \left(q_{\text{gg}}+i\right) \Gamma _g \\
 -\frac{1}{2} \left(q_{\text{gg}}+i\right) \Gamma _g & \text{$\delta
   $S}_g -\frac{i \Gamma _g}{2}+\delta _g
\end{array} \right],
\ee
\be
 \mathbf{\Delta}_{e}=\left[
\begin{array}{cc}
 \Delta -\frac{i \Gamma _e}{2}+\text{$\delta $S}_e & -\frac{1}{2} \left(q_{\text{ee}}+i\right) \Gamma _e
   \\
 -\frac{1}{2} \left(q_{\text{ee}}+i\right) \Gamma _e & \Delta+\text{$\delta $S}_e -\frac{i \Gamma _e}{2}+\delta
   _e \\
\end{array}
\right],
\ee
\ese
which account for the non-degeneracy by the energy shifts between the bound states $\delta_g$ and $\delta_e.$

\begin{figure}[tb]
\bt{cc}
\includegraphics[width=0.85\columnwidth]{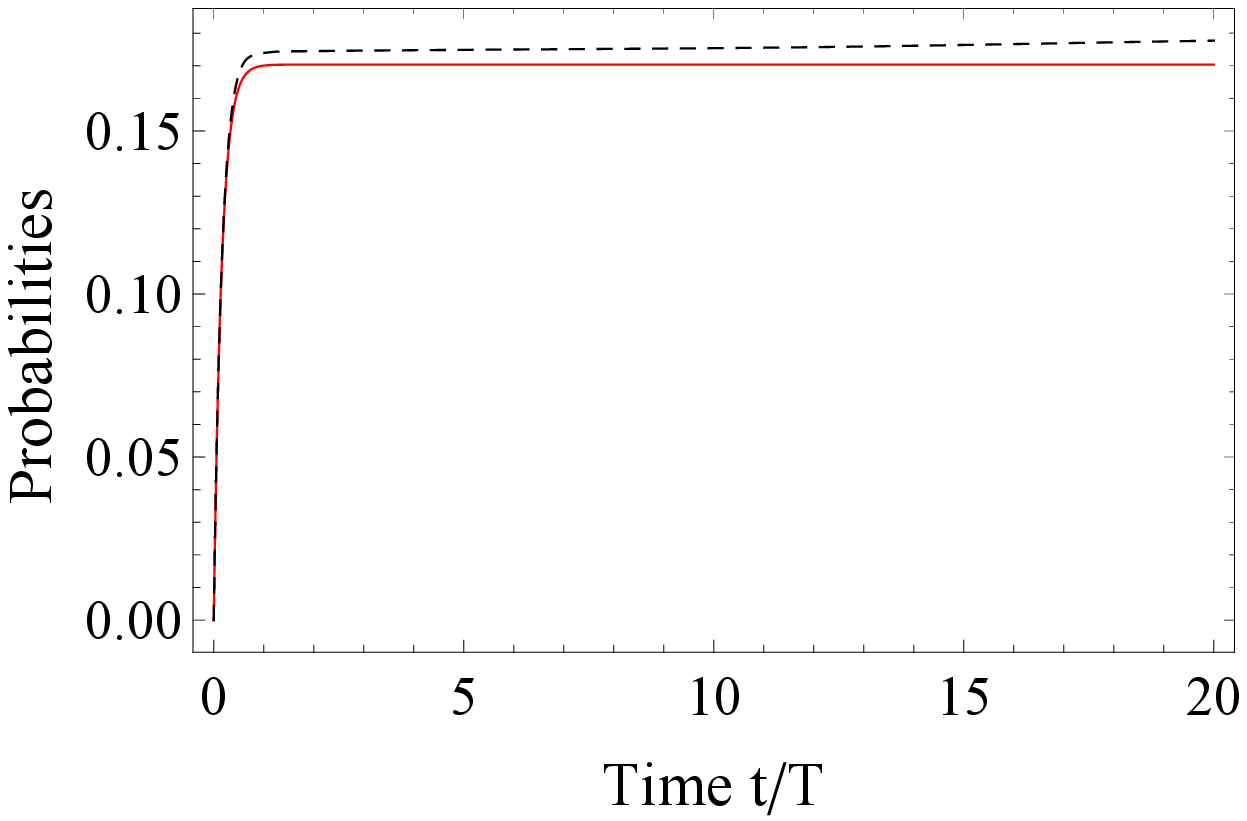}\llap{
  \parbox[b]{4.5in}{(a)\\\rule{0ex}{1.51in}
  }} \\
\includegraphics[width=0.85\columnwidth]{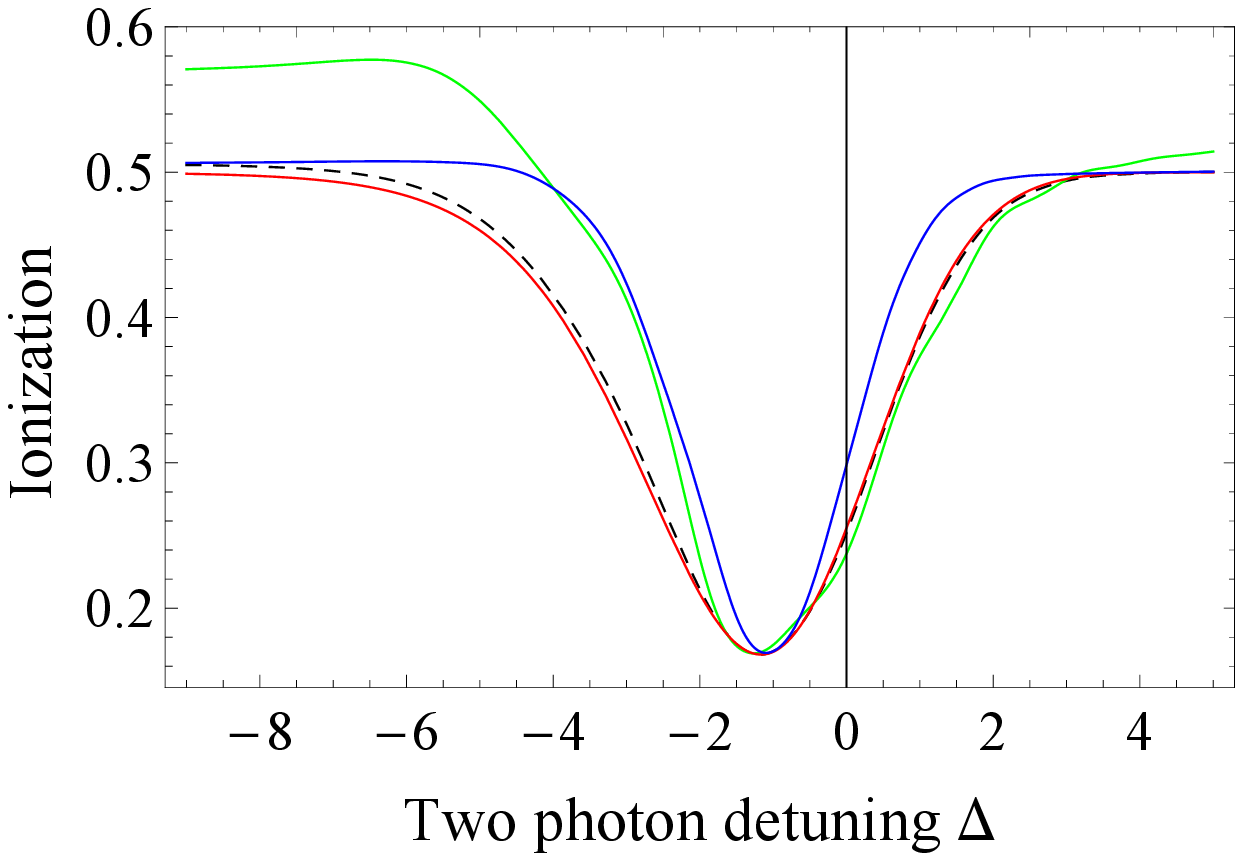}\llap{
  \parbox[b]{4.5in}{(b)\\\rule{0ex}{1.661in}
  }}
\et
\caption{(Color online)
(a) Ionisation as a function of the normalized time for the degenerate Hamiltonian (red solid line) of Eq.~\eqref{H0H1} and the non-degenerate (black dashed line) of
Eq.~\eqref{H_nd} for the system initialised in $c_{g_1}(0)=1$.
The excitation parameters are set to $\Gamma _g = 1.08 T^{-1},\Gamma _e = 2.09 T^{-1}, \delta S_g = 0.33 T^{-1},\delta S_e = 0.26T^{-1},
\delta_g = \delta_e = 0.2T^{-1}, q_{gg} = 2.3, q_{eg} = 2.4, q_{ee} = 2.5$.
(b) Fano profiles for the degenerate and non-degenerate  systems. The red line is the degenerate profile, while the blacked dashed line($\delta = 0.02T^{-1}$), the green($\delta = 0.2 T^{-1}$) and the blue($\delta = 0.02T^{-1}$) solid lines are the non-degenerate profiles calculated at $t/T = 10,$ except the blue line which is at $t/T = 20.$
}\label{fig:NondegSys}
\end{figure}

The effect of the non-degeneracy over time, depicted in Fig.~\ref{fig:NondegSys}(a), is to gradually increase the ionisation and deplete the bounded states,
thus to destroy the trapping. The effect of the degeneracy diminishes as the ratio of $\delta_k/\Gamma_{ij},$ which also regulates the width of the Fano profile[see Fig.~\ref{fig:NondegSys}(b)]. For large ratio the additional couplings proportional to the energy shifts become significant and destroy the bright and dark states picture. As evident[green line of Fig.~\ref{fig:NondegSys}(b)] these couplings can allow a higher amount of the population to be ionised when the trapping condition is not met. The minima of the Fano profiles however do not shift significantly, although the width changes at larger times[blue line].
Overall for reasonable time scales and large enough ionisation couplings the system can be treated as degenerate.
We note, however, that due to the nonlinear dependence of the system's response to the ionisation couplings such calculation for the validity of the degenerate
treatment should always be carried out in order to determine the ionisation strength and time window over which it is valid, as well as the error in the probabilities due to the approximation of degeneracy.

\section{Discussion and Summary}\label{Sec:conclusion}

In this paper, we explored the multistate LICS process consisting of two ground and two excited degenerate bound states coupled through a common continuum.
We reduced the dynamics of the system to a block-diagonal form by a rotation, mapping the evolution to bright and dark sub-systems.
This reduction of the complexity allowed us to derive separate trapping conditions for the sub-systems, in our example only for the bright Hamiltonian of
Eq.~\eqref{HuShr} since the dark states remain uncoupled. The assumption of degeneracy holds well as long as the ionisation strength is large enough compared to the energy shifts between the bound states.
Furthermore, we showed that the Fano profiles strongly depend on the initialisation of the system.
Initially populated bright state reproduces a standard ionisation profile, while initialisation in one of the ground states in the original basis sets an upper bound of the
ionisation of $\frac12$ because half of the population is trapped in the dark state.
The latter feature of the system can be an indicator of the number of states involved in the interaction and thus probes the structure of the system.
Another important application of the multistate LICS can be the generation of coherent  superpositions of Rydberg ions.
Naturally the Rydberg states are lying close to the continuum and can serve as the excited states in our model.
Thus a sample of Rydberg atoms initially prepared in a coherent superposition of ground states can be mapped through the continuum to an excited state.

\acknowledgments
KNZ acknowledges support from the project MSPLICS - P. Beron Grant from The Bulgarian National Science Fund (BNSF).


\end{document}